# Integration of diamond nanobeams with SnVs on $Al_2O_3$ waveguides for scalable quantum photonic chip application

Yeting Yang[1,2]*, Ryota Kitagawa[1], Tetsuya Miyatake[1], Masaharu Hida[1], Naoki Fushimi[1], Koki Kaminaka[1], Takuto Yamaguchi[1], Toshiki Iwai[1], Itsuki Takagi[1], Hidetsugu Matsukiyo[2], Satomi Ishida[2], Satoshi Iwamoto[2], Manabu Ohtomo[1], Toshiyuki Miyazawa[1], Kenichi Kawaguchi[1], Ryoichi Ishihara[3], Shintaro Sato[1]

[1] *Fujitsu Limited, 10-1 Morinosato-Wakamiya, Atsugi, Kanagawa 243–0197*

[2] *Research Center for Advanced Science and Technology, University of Tokyo 4-6-1 Komaba, Meguro-ku, Tokyo 153-8904*

[3] *Delft University of Technology, Mekelweg 5, Delft, The Netherlands 2628 CD*

*E-mail: yang.yeting@fujitsu.com

***Abstract*** *Tin-vacancy (SnV) centers in diamond are promising solid-state qubits for integrated quantum photonics. Here, we fabricate and characterize a diamond-on-$Al_2O_3$ dual-taper waveguide structure containing SnV centers, demonstrating optical coupling between the diamond nanobeam and the underlying $Al_2O_3$ waveguide. The devices are realized using a bilayer fabrication approach compatible with wafer-scale lithography. Clear guided $SnV^-$ emission is observed in all optically active devices, indicating effective optical coupling in the integrated structure. These results demonstrate a scalable fabrication approach toward integrating diamond color centers with photonic waveguides.*

Diamond color centers, including nitrogen-vacancy (NV) and group-IV vacancy centers (SiV, GeV, and SnV), are promising solid-state qubits owing to their long spin coherence times and optical addressability. SnV centers exhibit a large zero-phonon-line (ZPL) fraction, narrow optical linewidths, and high spectral stability, and can operate at elevated temperatures without the need for dilution refrigeration, making them attractive for integrated quantum photonic applications [1–6].

Realizing large-scale diamond-based quantum photonic systems requires efficient and low-loss on-chip optical interconnects to link spatially separated emitters. However, the available size of high-quality single-crystal diamond substrates is limited, making it difficult to realize large-scale photonic circuits. Hybrid integration of diamond nanostructures containing color centers with photonic waveguides offers a viable route to such architectures [1]. Aluminum oxide ($Al_2O_3$) is an appealing waveguide platform due to its wide bandgap, low optical loss, transparency at visible wavelengths, and compatibility with wafer-level nanofabrication processes [7, 8]. However, efficient optical interfacing between high-index diamond nanostructures and lower-index waveguides remains challenging due to mode mismatch and multilayer fabrication constraints.

The integration of diamond nanostructures with photonic waveguides using techniques such as transfer printing and pick-and-place has been demonstrated [9–11]. However, these approaches are typically based on serial processes, which may limit scalability. Therefore, lithography-compatible and wafer-scale integration approaches are highly desirable.

To address the need for scalable integration, we implement a wafer-compatible bilayer process to fabricate and characterize a vertically integrated diamond-on-$Al_2O_3$ photonic platform with a dual-taper region for efficient mode transfer between diamond nanobeams and underlying waveguides. The structure enables bidirectional waveguide-based excitation and collection of SnV emission: green excitation light is injected from the waveguide into the diamond nanobeam, and the resulting SnV emission is coupled back into the waveguide. Room-temperature optical measurements confirm that all optically active devices exhibit clear guided $SnV^-$ emission, demonstrating efficient optical coupling and propagation within the integrated diamond–$Al_2O_3$ structure. Statistical characterization across devices further highlights the potential of this lithography-based integration approach for large-area diamond quantum photonic circuits.

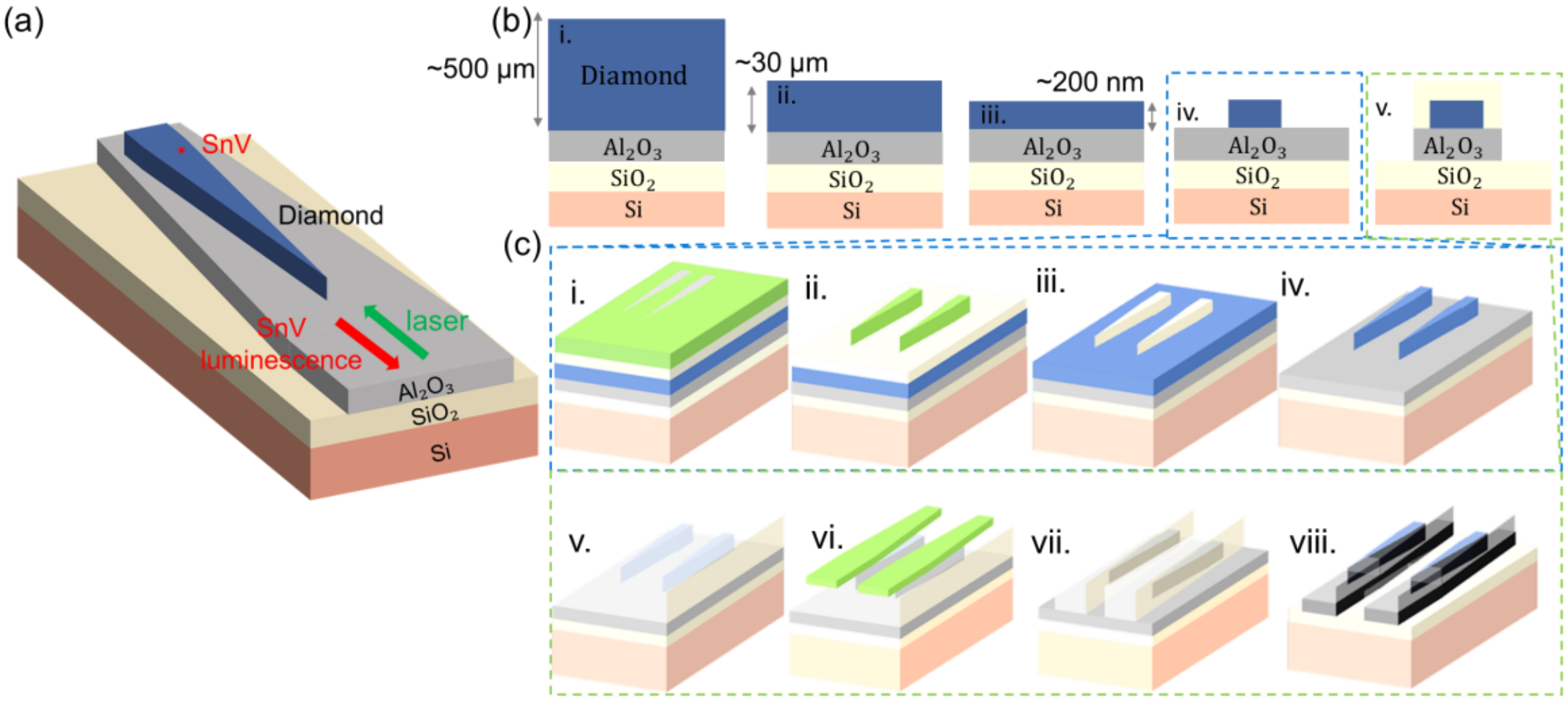

**Fig. 1. Concept and fabrication flow of the diamond–$Al_2O_3$ photonic device.**
(a) Schematic of the dual-taper structure, in which a diamond nanobeam containing SnV

centers is vertically integrated with an $Al_2O_3$ waveguide.
(b) Overview of the fabrication sequence: (i) wafer bonding; (ii) laser slicing and mechanical polishing; (iii) dry etching for thickness reduction; (iv) diamond nanobeam fabrication; (v) waveguide fabrication with hard-mask remaining.
(c) Detailed two-step etching process for forming the dual-taper structure: (i) $SiO_2$ deposition and EBL to define the nanobeam pattern; (ii) development; (iii) $SiO_2$ etching to transfer the pattern into the hard mask; (iv) vertical diamond etching to form the nanobeam; (v) $SiO_2$ deposition and resist spin-coating; (vi) EBL and development to define the waveguide pattern; (vii) $SiO_2$ etching to transfer the pattern into the hard mask; (viii) vertical $Al_2O_3$ etching to form the waveguide with hard-mask cladding.

As illustrated in Fig. 1 (a), the dual-taper structure consists of a 200 nm-thick diamond nanobeam aligned on top of a 300 nm-thick $Al_2O_3$ waveguide supported by $SiO_2$ on Si substrate. The $Al_2O_3$ thickness was chosen to ensure single-mode operation while preserving fabrication compatibility, based on the low-loss waveguide design in Ref. [7]. The 200 nm-thick diamond layer provides sufficient optical confinement, enabling preferential localization of the fundamental mode for propagation within the diamond and suppressing higher-order modes.

To realize this layer structure, we implemented a multi-step fabrication process combining wafer bonding, diamond thinning, and aligned bilayer nanofabrication. Owing to the extreme hardness and chemical inertness of diamond [3, 12], together with the strict requirements for sub-micron thickness control and precise device alignment, the fabrication process poses significant challenges.

Prior to nanofabrication, SnV centers were created by Sn ion implantation at a dose of $2 \times 10^{11}$ $cm^{-2}$ and an implantation energy of 350 keV, followed by annealing in an Ar atmosphere at 1500 °C for 2 h to activate the color centers [13–15].

An overview of the fabrication flow is shown in Fig. 1 (b). In Fig. 1 (b) (i), a 4 mm × 4 mm × 0.5 mm (100)-oriented single-crystal diamond substrate (Element Six, electronic grade) containing implanted SnV centers was bonded onto an $Al_2O_3$-on-$SiO_2$/Si substrate using a surface-activated direct bonding technique [16]. The bonded interface exhibited a shear strength exceeding 14 MPa, ensuring sufficient mechanical stability for subsequent processing.

As illustrated in Fig. 1 (b) (ii), the diamond thickness was first reduced to below 50 µm by YAG laser slicing. Mechanical polishing was then used to remove the laser-induced damage layer while further reducing the diamond thickness to ~30 µm. The polished surface exhibited a root-mean-square roughness below 0.2 nm and a peak-to-valley flatness better than 1.5 μm over a 10 μm × 10 μm area.

For step (iii) in Fig. 1 (b), the diamond layer was further thinned to approximately 200 nm using reactive ion etching (RIE) to enable efficient coupling of photons emitted from SnV centers in the diamond nanobeam into the $Al_2O_3$ waveguide [17–23]. A pre-clean step employing Ar/$CF_4$ plasma was introduced to suppress micro-mask formation [12, 24] arising from particle redeposition during $O_2$ plasma etching. Subsequent $O_2$-based RIE thinning was performed at an etch rate of 117 nm/min.

The diamond nanobeam [Fig. 1 (b) (iv)] and the $Al_2O_3$ waveguides [Fig. 1 (b) (v)] were fabricated in aligned position to form the hybrid structure.

In Fig. 1 (c) (i), the shape and position of the diamond nanobeam were defined by electron beam lithography (EBL) using the negative resist. After development [Fig. 1 (c) (ii)] a 150 nm-thick $SiO_2$ hard mask deposited via chemical vapor deposition on the diamond layer was etched using $CHF_3$-based RIE with an inductively coupled plasma (ICP) power of 125 W and a bias power of 100 W to transfer the pattern [Fig. 1 (c) (iii)].

The exposed diamond regions were then etched using $O_2$ plasma [25, 26] with an ICP power of 50 W and a bias power of 200 W [Fig. 1 (c) (iv)], achieving an etch rate of 20.8 nm/min with a total etch time of 12 min, including an over-etch to ensure full isolation of the 200 nm-thick diamond layer. The use of low ICP power effectively suppressed micro-mask formation and minimized damage to the underlying $Al_2O_3$ layer.

Following the formation of the diamond nanobeam, a 600 nm-thick $SiO_2$ hard mask was deposited and spin-coated with EB resist for waveguide fabrication [Fig. 1(c) (v)]. The waveguide pattern was defined by EBL and development [Fig. 1 (c) (vi)] and then transferred to the $SiO_2$ mask via $CHF_3$-based dry etching at an RF power of 100 W [Fig. 1 (c) (vii)].

Finally, the $Al_2O_3$ layer was etched using a $Cl_2$/$BCl_3$ plasma [7, 8, 27] with an RF power of 200 W and a bias power of 100 W to form the waveguide with the hard-mask cladding [Fig. 1 (c) (viii)], resulting in an etch rate of 35.0 nm/min for $Al_2O_3$ and leaving a 300 nm-thick $SiO_2$ cladding on the device.

Scanning electron microscope (SEM) images of the fabricated devices are shown in Fig. 2. Fig. 2 (a) presents a tilted-view SEM image of the diamond nanobeams after the diamond etching step, showing an array of nanobeams fabricated on the chip. The inset displays a magnified view of a representative nanobeam, clearly revealing the tapered region with a sharp tip. The taper is designed with a sharp triangular profile with an angle of ~2.86°, while the fabricated structure exhibits a minimum tip width of ~20 nm, which is reasonable considering EBL resolution.

Fig. 2 (b) shows the SEM image after completion of the $Al_2O_3$ waveguide fabrication, corresponding to the final bilayer structure. The regions with and without diamond nanobeams can be clearly distinguished, confirming the successful realization of the vertically integrated diamond–$Al_2O_3$ waveguide structure. Note that the remaining $SiO_2$ hard mask layer (~300 nm thick) clads the diamond nanobeam, making its contour difficult to resolve in the SEM image.

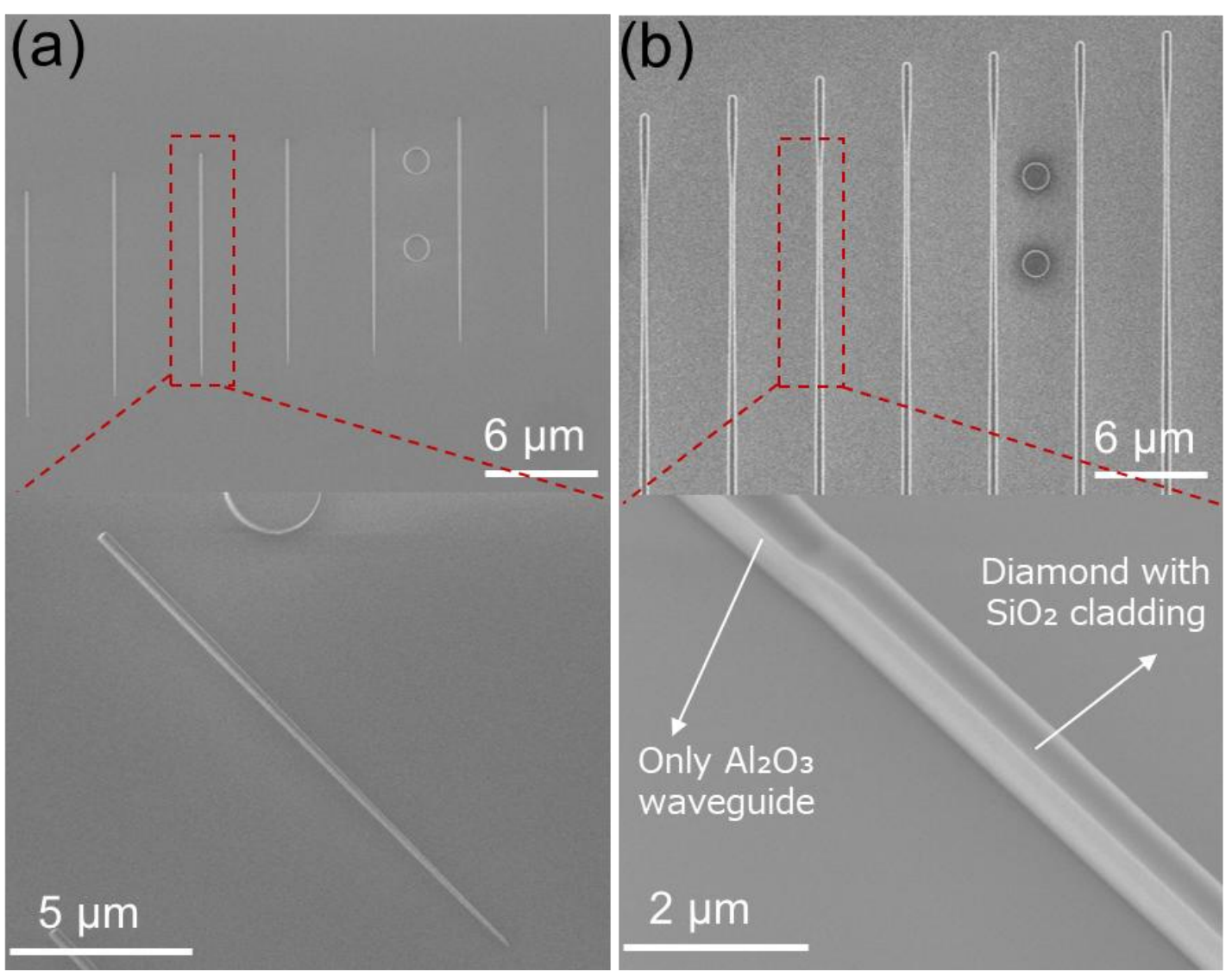

**Fig. 2. Structural characterization of fabricated diamond-on-waveguide devices.**
(a) Tilted-view (20°) SEM image of the fabricated diamond nanobeams after the diamond etching step (Fig. 1 (c) (iv)), showing an array of seven nanobeams. The inset presents a magnified SEM image (rotated-45°) of a representative nanobeam.
(b) SEM image of the completed diamond-on-waveguide device after the waveguide fabrication step (Fig. 1 (c) (viii)). The inset shows a magnified view of a representative structure. Arrows indicate the regions corresponding to the bare $Al_2O_3$ waveguide and the waveguide supporting the diamond nanobeam.

For the hybrid structure, room-temperature optical characterization was performed using a reflection-type configuration through waveguide, as schematically illustrated in the inset of Fig. 3 (a). A 520 nm laser was coupled into the $Al_2O_3$ waveguide through a cleaved facet via a lensed fiber. The excitation light propagated to the diamond nanobeam to excite the SnV centers, and the emitted photons were collected in the counter-propagating direction through the same waveguide port and directed to a spectrometer. A 600 nm long-pass filter was used to block the excitation laser, which slightly attenuates PL signals in the 600–602 nm region, without affecting the evaluation of the $SnV^-$ emission.

We observed representative photoluminescence (PL) spectra, as shown in Fig. 3 (a). The red and black curves correspond to PL collected from a waveguide coupled to a diamond nanobeam and from a reference waveguide without a nanobeam, respectively. Negligible PL is observed from the reference waveguide. In contrast, the bilayer device exhibits a pronounced peak at ~620 nm, corresponding to the $SnV^-$ ZPL emission [13–15]. In addition to the main ZPL peak, weaker emission features are observed around 650 nm and 660 nm, which have been previously reported in Sn-implanted diamond and attributed to other Sn-related defect configurations or sideband emissions [27–30]. Based on the implantation dose and the estimated activation yield of $SnV^-$ centers, each nanobeam is expected to host several tens of emitters, consistent with the observed ensemble emission profile. The measured ZPL linewidth is ~6 nm, which can be attributed to phonon-induced broadening at room temperature and strain inhomogeneity, consistent with previously reported values [14]. These results confirm optical coupling between the $Al_2O_3$ waveguide and the diamond nanobeam for both excitation and $SnV^-$ emission.

To quantitatively evaluate the precision of the fabrication process, a statistical analysis is performed for 216 waveguide-coupled devices. A key advantage of the bilayer fabrication method developed in this work is its scalability and compatibility with wafer-scale lithographic processing. On the sample chip used in this study, a total of 216 waveguide-coupled devices were fabricated and systematically characterized. Fig. 3 (b) presents a heatmap of the PL spectra collected from all devices. The devices are grouped into eight categories according to the diamond nanobeam geometry, with 27 nominally identical devices in each group. We observed clear $SnV^-$ emission in the majority of devices. A device is classified as "$SnV^-$ observed" if its PL peak intensity at 620 nm exceeds one-third of the peak intensity for the brightest device within the corresponding size group. Based on this criterion, 179 out of 216 devices exhibited detectable $SnV^-$ emission, corresponding to an overall optical yield of 82.9%.

To investigate the origin of the 37 devices without detectable $SnV^-$ emission, we analyzed each device in detail shows in Fig. 3 (b). The possible causes were categorized into two types: waveguide-related issues "WG issue" and diamond-related issues "Diamond issue". WG issues include defects in the $Al_2O_3$ waveguide, such as discontinuities, pattern deformation, incomplete dual-taper formation, and scattering induced by surface particulates or voids in the $SiO_2$ underlayer. Diamond issues refer to defects in the diamond nanobeam itself, including structural damage, surface roughness, or absence/misalignment of active $SnV^-$ centers. All 37 devices are found to exhibit WG issues, while no diamond issues are observed, indicating that the diamond nanobeams remained intact and that non-emission arose entirely from waveguide fabrication imperfections. The primary failure mechanism is identified as waveguide discontinuity, caused in 15 devices by diamond micro-pillar formation resulting from residual micro-masking during the diamond thinning process.

Fig. 3 (c) summarizes the collected PL intensity as a function of diamond nanobeam area. Two geometries were fabricated: pentagon-shaped nanobeams with both rectangular and triangular tapers, and triangular nanobeams with only a triangular taper. Within each geometry, the PL intensity increases approximately linearly with area, suggesting that optical excitation and emission propagate throughout the entire nanobeam. A slight reduction in PL intensity is observed for the largest pentagon devices, which may indicate structural imperfections, although the dominant mechanism remains unclear. Pentagon-shaped devices show higher PL intensity than triangular ones of the same area, likely due to stronger emission from their central rectangular region, which is located away from etched edges and may benefit from more effective light coupling.

Overall, these results demonstrate that the dual-taper bilayer platform enables efficient optical interfacing between diamond color centers and integrated waveguides. The fabrication process is compatible with wafer-scale lithography and supports parallel fabrication of large device arrays, providing a scalable route toward integrated diamond quantum photonic circuits.

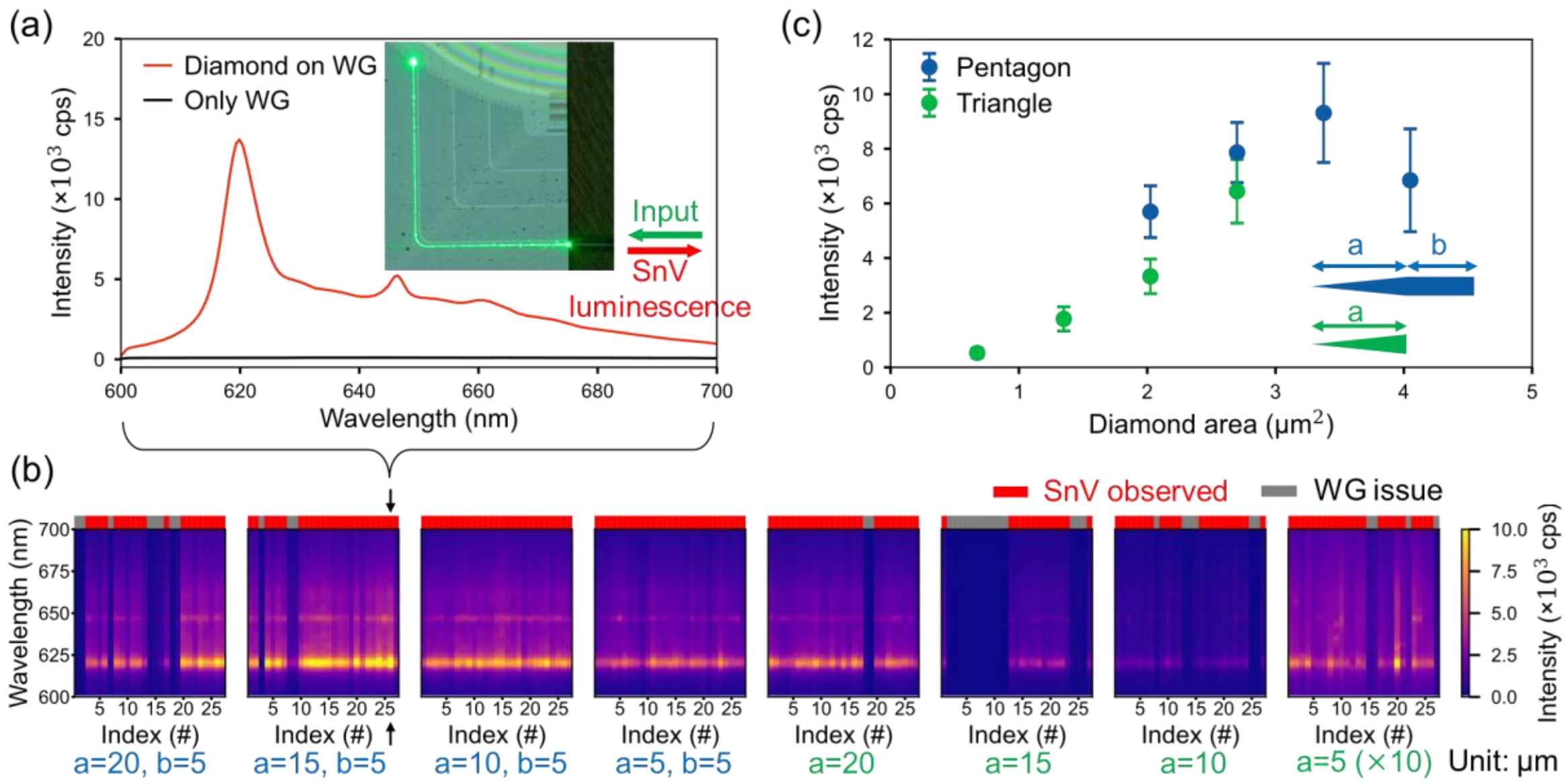

**Fig. 3. Optical characterization and statistical device yield.**
(a) Photoluminescence spectrum of a diamond–on–$Al_2O_3$ waveguide device, showing the $SnV^-$ ZPL emission coupled into the waveguide. Inset: Optical images under 520-nm excitation. (b) Heatmap of PL spectra collected from multiple devices across the chip, illustrating high-yield $SnV^-$ emission. Devices are grouped by diamond geometry: pentagon-shaped nanobeam with triangular taper (green) and triangular taper only (blue), defined by the triangular taper length *a* and rectangular section length *b*. For each geometry, 27 devices were measured. "WG issue" represents devices with no detectable PL. (c) Statistical correlation between collected PL intensity and diamond area for the two device geometries.

In conclusion, we demonstrate a fully lithographic and vertically integrated diamond–on–$Al_2O_3$ waveguide structure incorporating a dual-taper architecture for optical coupling of diamond color centers. The devices are fabricated using a wafer-scale bilayer process, enabling the realization of large device arrays. Room-temperature photoluminescence measurements confirm that $SnV^-$ emission from the diamond nanobeam is coupled into the underlying waveguide, allowing both optical excitation and photon collection through a single waveguide port.

Statistical characterization of 216 devices yields an overall optical yield of 82.9%. Systematic variation of diamond geometry and nanobeam area shows a monotonic increase in photoluminescence intensity consistent with an increasing number of optically active emitters. On a single chip, more than 100 waveguide-coupled emitters are observed, demonstrating the scalability of the platform for integrated diamond quantum photonics.

Acknowledgment: This work was partially supported by "Advanced Research Infrastructure for Materials and Nanotechnology in Japan (ARIM)" of the Ministry of Education, Culture, Sports, Science and Technology (MEXT). Proposal Number JPMXP1225NM0096.